\documentclass{aa}  
\usepackage{graphicx}
\usepackage{natbib}
\usepackage{txfonts}
\newcommand{\ROSAT}{{\it ROSAT}}
\begin{document}
   \title{Warm-hot intergalactic medium contribution to baryonic matter}

   \author{A. M. So\l tan
          \inst{}
          }

   \offprints{A. M. So\l tan}

   \institute{Nicolaus Copernicus Astronomical Center,
              Bartycka 18, 00-716 Warsaw, Poland\\
              \email{soltan@camk.edu.pl}
             }

   \date{Received ; accepted }

 
  \abstract
   {Hydrodynamical simulations indicate that substantial fraction of baryons
    in the Universe remains in a diffuse component -- Warm-Hot Intergalactic
    Medium (WHIM).}
   {To determine physical properties (spatial distribution, temperature and
    density) of WHIM.}
   {Spatial structure of the soft extended X-ray emission surrounding field
    galaxies is carefully investigated using the XMM-Newton EPIC/MOS observations.
    Angular correlations between the galaxy distribution and
    the soft X-ray background extending over several arcmin result from the large
    scale clustering of galaxies. At small angular scales (below $\sim 2\arcmin$)
    the excess of the soft flux is interpreted as the genuine emission from
    halos of the WHIM surrounding individual galaxies.}
   {Bulk parameters of the average WHIM halos associated with galaxies in
    the sample are estimated. Typical halo has a characteristic radius of
    $\sim 0.3$\,Mpc and a mass of $4 - 7\cdot 10^{11}$\,M$_{\sun}$. The average
    density of the WHIM in the local universe amounts to
    $7 - 11\cdot 10^{-32}$\,g\,cm$^{-3}$ ($\Omega_{\rm WHIM} = 0.7 - 1.2$\,\%).}
   {Observations of the X-ray WHIM emission are in good agreement with the numerical
    simulations, but accuracy of the observational material is insufficient to
    constrain the theory of WHIM. A series of deep observations of a moderately
    numerous sample of low redshift galaxies with high resolution instruments
    of $ Chandra$ would significantly improve our estimates of the WHIM
    parameters.}

   \maketitle
%

\section{Introduction}

Both theoretical arguments and observational data point out that a major
fraction of baryons in the local universe remains in the form of the
diffuse component. Following the first hydrodynamical simulations by
\cite{cen99}, several independent groups (e.g. \citealt{dave01, bryan01,
croft01}) investigated the evolution of the primordial gas distributed
outside large concentrations of mass, viz. galaxies and clusters.
According to simulations $30-40$\,\% of the baryonic matter has not yet
concentrated in gravitationally bound objects. This diffuse component
gradually flows toward potential wells created by (mostly) non-baryonic
dark matter.

A rate at which the gas flows toward the galaxies depends on the structure of
the gravitational potential and the efficiency of the non-gravitational
heating, the so-called feedback (e.g. \citealt{zhang02}). This comprehensive
term stands for complex processes associated with transfer of energy and matter
from galaxies back to the intergalactic medium.
Although details of these processes are not fully recognized and modeled, basic
conclusions drawn from the calculations by all the research groups agree. It is
found that the interaction between the infalling gas and the galaxy generates
shocks which heat the intergalactic material. In effect galaxies
are surrounded by halos of plasma with temperatures and densities
distinctly higher that in the areas not affected by the feedback.

Simulations predict a wide range of the intergalactic plasma temperatures.
In the local universe characteristic temperatures of the model gas accumulating
around galaxies increase to $10^5 - 10^7$\,K, while in the more
remote regions gas remains at the temperatures of $\sim 10^4$\,K.
Also the densities of the heated gas fluctuate and span a wide range of
magnitudes. A substantial fraction of baryons is predicted to
undergo rather moderate compression with a median overdensity
of $\sim 20$ over the cosmic mean (\citealt{dave01}). However,
most of the X-ray--emitting gas has densities $100 - 1000$ times greater
than the mean (\citealt{croft01}). In this way,
the cosmological simulations of the matter evolution
indicate that a measurable fraction of baryons constitutes a specific
{\it phase} of matter, so-called {\it Warm-hot intergalactic medium}
-- WHIM (\citealt{dave01}).

A cloud of the WHIM might reveal its existence either through the intrinsic
emission of the thermal radiation or through absorption lines induced in
spectra of the background objects.  Potentially both effects can be used to
evaluate physical parameters of the gas (spatial structure, temperature,
density) and eventually to constrain models of the WHIM evolution. Estimates
based on the simulations indicate that expected equivalent width of absorption
lines as well as luminosities generated by the individual WHIM clouds are
extremely low. Nevertheless, detections of absorption lines produced by the
intergalactic hot plasma have been reported by several groups
(e.g.~\citealt{tripp00, nicastro05}, and references therein).  Observations of
O$^{6+}$ K$\alpha$ absorption lines in the spectrum of the blazar MKN 421 by
\cite{nicastro05} have been used to make a statement on statistical
characteristics of the WHIM and its contribution to the local cosmological mass
density.

Extragalactic thermal radiation has been searched by \cite{kuntz01}. These
authors investigated the surface brightness of the soft XRB in the \ROSAT\
All-Sky Survey (RASS) and  detected a smooth thermal component with
$k{\rm T} = 0.23$\,keV which most likely was a mixture of WHIM and the
Galactic halo emission. Existence of gaseous halos of a radius $30 - 40$\,kpc
surrounding the spiral galaxy NGC\,$5746$ has been reported recently
by \cite{pedersen05}. The authors argue that the observed halo is formed
from the in-flowing gas rather than the matter expelled in the supernova
explosions.

In the present paper we continue our WHIM studies by investigating the soft
X-ray emission generated by this phase of the baryonic matter.  The main
objective of the paper is  to determine physical characteristics of the
emitting plasma and -- eventually -- to assess contribution of the WHIM to the
total mean space density of the baryonic matter. In a series of earlier papers
we have analyzed the structure of the soft X-ray background (XRB). Using the
RASS maps we have detected soft enhancements of the XRB flux
around clusters of galaxies (\citealt{soltan96}) and galaxies
(\citealt{soltan97}). Assuming a thermal origin of this excess flux,
\cite{soltan02} have shown that temperature of the emission $k{\rm T} \la
0.5$\,keV in agreement with the expected temperature of the WHIM. We have
extended our analysis toward smaller angular (and spacial) scales using the
\ROSAT\ and XMM-Newton pointing observations and confirmed that the surface
brightness of the soft X-ray excess flux increases as the distance to the
galaxy diminishes, again as expected for the WHIM structure
(\citealt{soltan05}).

In this paper spatial properties of the WHIM is investigated in greater details.
Scrupulous analysis of the XRB structure in the XMM-Newton data allowed us to
determine the characteristic size and luminosity of the X-rays emitting WHIM
halo around field galaxies.

Two fundamental factors complicate observations of the WHIM emission. First,
the weak and diffuse WHIM signal is observed against the highly variable
background produced by a whole variety of discrete sources. The integral
background is dominated by point-like sources identified mostly
with all kinds of AGN (e.g. \citealt{alexander03}, \citealt{worsley05}, and
references therein), leaving not much space for the diffuse component.
To increase signal-to-noise ratio of the WHIM component all the detected
point sources are removed from the data. Still, the sources below
the detection threshold generate fluctuations which impede the WHIM search.
Second, due to relatively low temperatures, the thermal emission by the WHIM
is expected to be soft, typically below $1$\,keV, with most of the flux
at $\sim 0.5$\,keV. At such low energies the absorption of extragalactic photons
by the cold gas in the Galaxy is significant and the search for the WHIM emission
has to be limited to high galactic latitudes.

In the next section a short description of the observational material used in the
analysis is given. The method of calculations and the raw results are presented
in Sect.~\ref{method}. The interpretation of the detected signal is performed
in Sect.~\ref{results} and \ref{densities}. Brief discussion is given in
Sect.~\ref{discussion}.

\section{Observational material \label{observations}}

All the simulations of the WHIM emission (e.g. \citealt{bryan01},
\citealt{croft01}) and our earlier estimates show that the surface brightness
of an individual WHIM cloud does not exceed few percent of the average
XRB level. To achieve meaningful signal-to-noise ratio for the WHIM emission
one requires extensive set of data. An effective method to measure the  weak
enhancements of the XRB around galaxies is to calculate the cross-correlation
function (CCF) of the XRB and the sample of galaxies. To draw quantitative
conclusions from the CCF, one has to remove from the XRB maps systematic effects
which could mimic the diffuse signal. Also the sample of galaxies should have
well-defined statistical properties.

\subsection{EPIC MOS data}

In the present analysis the X-ray data are extracted from the public
archive\footnote{XMM-Newton Science Archive:\\
http://xmm.vilspa.esa.es/external/xmm\_data\_acc/xsa/index.shtml} 
of the XMM-Newton EPIC/MOS observations. The pointings are selected in the
same way as in the work by \cite{soltan05} and for the full description of the
data reduction the reader is referred to the Sect.\,2.2 of that paper. Here
only the main points are recalled and some modifications of the original procedures
described.

All the data obtained with the MOS1 and MOS2 detectors in a ``Full Frame'' mode
with the thin filter have been inspected and only pointings ``suitable'' for
the present investigation have been used. The pointings with a strong source,
known extended source or the high particle background have been classifies as
``unsuitable''. To optimize our search of the WHIM signal, new (in comparison
to \cite{soltan05}) energy bands have been selected: $0.3-0.5$\,keV,
$0.7-1.0$\,keV, $1.0-1.35$\,keV, $1.9-3.0$\,keV and $3.0-4.5$\,keV.
Energy gaps $0.5-0.7$\,keV and  $1.35-1.9$\,keV diminish  Galactic contribution
and the strongest internal fluorescent lines, respectively
(\citealt{nevalainen05}).

It was pointed by \cite{pradas05}, that to study faint extended sources such as
the WHIM using the XMM-Newton EPIC detectors ``all systematic effects must
be well understood and reliable methods to eliminate their contribution have to
be developed''. In the present approach the WHIM emission is analyzed taking
advantage of a large number of available observations. Several hundred
pointings had been inspected and above $150$ were accepted for further
processing.  A median exposure time of the observation used in the
investigation is below $10$\,ks. It was not possible to investigate
instrumental effects reported by \cite{pradas05} for the each pointing
separately. However, to account for the possible irregularities of the CCD
performance, we have introduced a correction to the procedure described in
\cite{soltan05}. Separately for the MOS 1 and MOS 2 cameras and for the each of
our energy bands ``a sensitivity map'' has been generated. The sensitivity map
has been constructed as follows.  First, all the point-like sources found in the
individual observation have been removed. Then, the counts in the detector
coordinates for all the pointings have been added. In the same way ``integral''
exposure map has been created. Finally, the summed count distribution has been
divided by the exposure map. Since the resultant distribution of countrates in
pixels is based on a large number of pointings, fluctuations of the cosmic
signal are smoothed out. Thus, spatial variations of the countrates represent
the mean inhomogeneities of the detector sensitivity. Although this procedure
is unable to deal with the time dependent instrumental effects, it allows for
the coherent treatment of a large set of the EPIC MOS observations.

\subsection{Galaxy data}

The galaxy sample has been extracted from the APM Galaxy Survey
(\citealt{maddox90a}, \citealt{maddox90b}) using the NASA/IPAC Extragalactic
Database (NED). All galaxies with magnitudes between $17$ and $20$ have been
included into the investigation. This magnitude selection criterion is the same
as in the subsample of the APM Galaxy Survey discussed by \cite{maddox96}.  The
present galaxy sample, drawn from the statistically complete survey and with
well-defined magnitude limits, is more homogeneous than the sample used by
\cite{soltan05} and some statistical characteristics of the APM galaxies
determined by \cite{maddox96} are directly applicable to the present
investigation (see below).

\begin{figure}
\centering
\includegraphics[width=9cm]{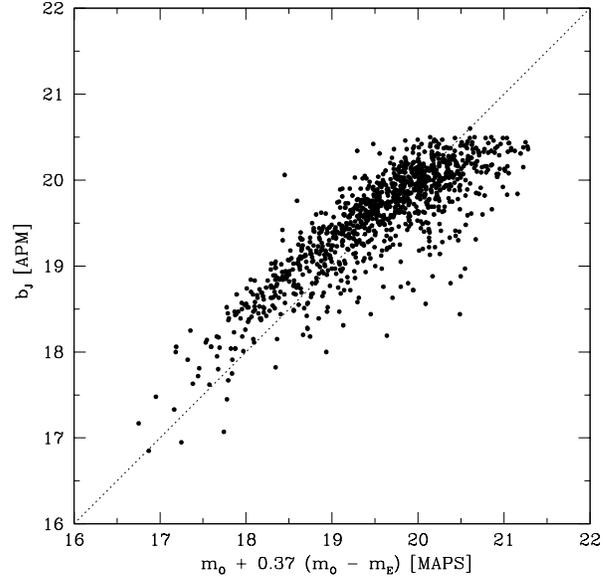}
\caption{Comparison of the original $b_J$ magnitudes from the APM Galaxy Survey
and the magnitudes derived from $m_O$ and $m_E$ in the MAPS Catalogue.}
\label{magnitudes}
\end{figure}

In the areas not covered by the APM Survey, galaxies extracted from the
Minnesota Automated Plate Scanner (MAPS)\footnote{The MAPS database is
supported by the University of Minnesota, available at http://aps.umn.edu/.}
Catalog of the POSS I have been used. Selection criteria applied to the the
MAPS Catalogue have been constructed to generate sample of objects with
statistical characteristics resembling the APM galaxy sample. Using more than
1100 galaxies common to the APM and MAPS samples, a linear transformation
between the O and E magnitudes from MAPS and $b_J$ have been found in the form:
$ b_J = m_O - 0.37 (m_O - m_E)$.  Fig.~\ref{magnitudes} shows the distribution
of magnitudes in both samples.  Although the correlation indicated by the
dotted line is not perfect, we expect that galaxies selected from the MAPS have
clustering properties and the redshift distribution similar to the APM
galaxies.

\section{Correlation analysis \label{method}}

The average intensity of the XRB at a distance $\theta$ from a randomly chosen
galaxy is given by the formula:

\begin{equation} \rho(\theta) = {\sum n_{\rm
cnt} \over \sum n_{\rm pxl} \cdot t_{\rm exp}}\,, \label{ccf_def}
\end{equation}
where the sums extend over all pointings and all galaxies, $n_{\rm cnt}$
denotes the total number of counts recorded in $n_{\rm pxl}$ pixels separated
by the angle $\theta$ from the galaxy, and $t_{\rm exp}$ denotes the
appropriate exposure time (see \cite{soltan05} for the detailed description).
All the data were binned into $4\arcsec \times 4\arcsec$ pixels. The total
amplitude of the countrate $\rho(\theta)$ obtained from the set of actual
observations is a mixture of the cosmic and local X-ray photons as well as
charged particles.  One should note, however, that any potential systematic
variations of the $\rho$ with the separation angle $\theta$ could result
exclusively from the genuine changes of the average extragalactic XRB flux
associated with the galaxy sample. Thus, the data on the absolute level
of the contaminating counts are not crucial for the determination
of the $\rho(\theta)$ slope.


The flux distribution around galaxies is related to the CCF, $w(\theta)$,
in a standard way:
\begin{equation}
w(\theta) = \rho(\theta)/\bar{\rho} - 1,
\end{equation}
where $\bar{\rho}$ denotes the total average countrate. The average counts
$\bar{\rho}$ and consequently the CCF amplitude are affected by the non-cosmic
signal.

\begin{figure}
\centering
\includegraphics[width=9cm]{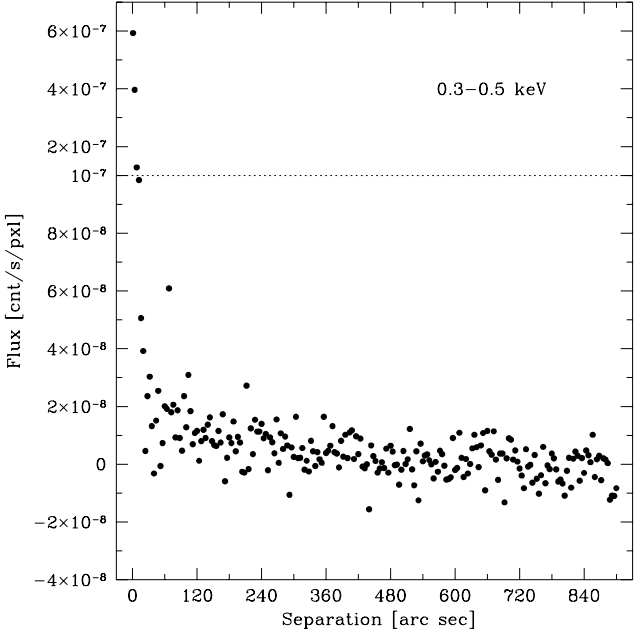}
 \caption{The countrate distribution in the soft energy band ($0.3-0.5$\,keV)
  vs. angular distance from the galaxy averaged over the galaxy sample.}
 \label{ccf_soft}
\end{figure}

\begin{figure}
\centering
\includegraphics[width=9cm]{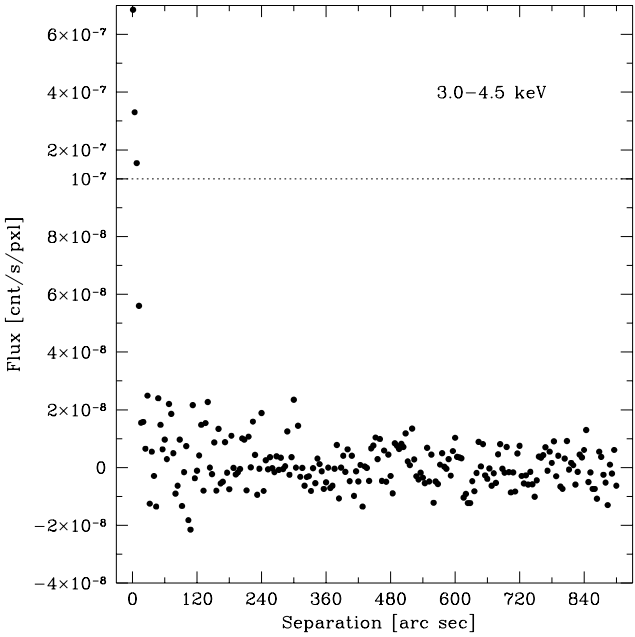}
 \caption{Same as Fig.~\ref{ccf_soft} in the hard energy band ($3.0-4.5$\,keV).}
 \label{ccf_hard}
\end{figure}

To isolate the XRB fraction correlated with the galaxy distribution
from the total counts and to remove residual instrumental effects in the
$\rho(\theta)$ distribution, a large number of random sets of ``galaxy'' samples
have been generated. The average $\rho_{\rm rand}(\theta)$ countrate
distribution of the randomized data provides an estimate of the XRB component
uncorrelated with the galaxy sample. The difference between both distributions:

\begin{equation}
\delta\rho(\theta) = \rho(\theta) - \rho_{\rm rand}(\theta),
\end{equation}
which represents the net correlated signal is shown for the soft
($0.3-0.5$\,keV) and hard ($3.0-4.5$\,keV) energy bands in Figs.~\ref{ccf_soft}
and \ref{ccf_hard}, respectively.  Note change of scale at the $y$-axis above
the dotted line in both figures. The distribution of countrates in the hard
energy band is noticeably flat at separations above $\sim\!\!1\arcmin$, while
in the soft band systematic decrease of the flux with the increasing distance is
present over a wide range of separations.  Although the same effect is visible
in Fig.~4 of \cite{soltan05}, the Fig.~\ref{ccf_soft} is shown here as it is
based on the statistically well-defined, complete sample of galaxies.  The
scatter of points in Fig.~\ref{ccf_soft} is now reduced due to larger number of
galaxies involved in the analysis.  Since the $\delta\rho(\theta)$ function is
closely related to the average X-ray emission associated with a single galaxy
(see below), the statistical completeness of the galaxy sample is crucial for
the estimates of the WHIM emission.

The average surface brightness excess of the XRB around the galaxy is produced
by the source(s) ``associated'' with the galaxy itself and by the nonuniform
distribution of neighbouring galaxies. So, the integral X-ray emission produced by
stellar sources within the galaxy and generated in the extended WHIM halo
surrounding the galaxy, is defined here as the source ``associated'' with the
galaxy. Let $s_i$ denotes the total average flux produced by the galaxy in the
$i$-th energy band, $w_{gg}(\theta)$ -- the galaxy ACF, and $n_g$ -- the
average concentration of the sample galaxies on the sky.  The excess XRB flux
above the background level is approximately given by the formula:

\begin{equation}
\delta\rho_i(\theta) \approx s_i\cdot f_i(\theta) +
                   s_i\cdot w_{gg}(\theta)\cdot n_g,
\label{delta_rho_1}
\end{equation}
where the $f_i(\theta)$ function describes the distribution of the galaxy
emission in $i$-th band convolved with the instrument Point Spread Function
(PSF).  
If the WHIM contribution to the $\delta\rho_i(\theta)$ function is negligible and
the angular size of the galaxy X-ray emission is much smaller then the width of the PSF,
we have:
\begin{equation} \delta\rho_i(\theta) =
s_i\cdot {\rm P}_i(\theta) + 
                    s_i\cdot w_{gg}(\theta)\cdot n_g,
\label{delta_rho_2}
\end{equation}
where ${\rm P}_i(\theta)$ is the PSF in the $i$-th band.
In this case both components of the excess flux at the right hand side of
Eq.~\ref{delta_rho_2} are easily identified in
in Fig.~\ref{ccf_soft}. A strong peak at small angular separations (below
$\sim 1\arcmin$) is generated by the point-like galaxy emission, while at larger
separations a smooth decrease of the $\rho_i(\theta)$ results from the declining
density of neighbouring galaxies.

The X-ray emission generated within the galaxy either due to the nuclear
activity or in the galactic X-ray sources (X-ray binaries, supernova remnants)
is strongly concentrated and in the present sample of galaxies angular extent
of the galaxy emission is smaller than width of the PSF.  On the other hand,
the genuine emission produced by the WHIM surrounding the galaxy at zero lag is
expected to be substantially more extended than the width of the PSF. One
should note that this component would be difficult to distinguish from the
smoothed flux originating in the neighbouring galaxies.

\subsection{The effective Point Spread Function \label{psf}}

\begin{figure}
\centering
\includegraphics[width=9cm]{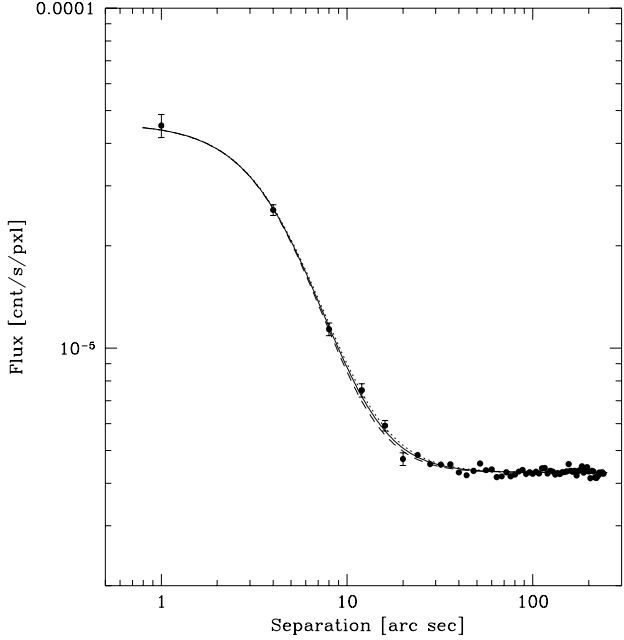}
\caption{The countrate distribution in the merged 5 energy bands vs. distance
from a quasar averaged over the distant quasar sample (points with error bars).
The curves represent fits of the King profile to the data: solid curve - 5
energy bands merged, dotted curve - 3 soft energy bands merged, dashed curve -
2 hard energy bands merged.} \label{qso_psf}
\end{figure}

Since the correlation flux $\delta\rho_i(\theta)$ is generated by sources
scattered over the entire field of view, the corresponding point source signal
should be averaged in the same way. It is assumed that a sample of
serendipitous quasars with redshifts greater than $0.3$ in the investigated
pointings generate countrates which mimic the shape of the point sources
contribution to the $\delta\rho_i(\theta)$.  The QSOs have been extracted from the
NED and the average countrate distribution around the sample objects has been
determined in the same way as for the galaxy sample. The correlation signal for
the quasar sample, $\rho_{\rm QSO}(\theta)$, is shown in Fig.~\ref{qso_psf}.
The data points representing the countrates summed over all 5 energy bands
are adequately approximated by the King function:
\begin{equation}
\rho_{\rm QSO}(\theta) = \rho_0 \cdot
          \left[ 1 + \left( {\theta \over \theta_0} \right)^2\right]^{-y} + \rho_b,
\label{rho_qso}
\end{equation}
with four simultaneously fitted parameters $\rho_0$, $\theta_0$, $y$ and
$\rho_b$. Although two shape parameters of the  PSF, viz. $\theta_0$ and $y$,
depend on the energy, the effect is neglected in the present analysis. This is
because the PSF width variations over our energy bands are minute and do not
affect noticeably the present calculations. It is demonstrated in
Fig.~\ref{qso_psf}, where three sets of the shape parameters are used to fit
the observed countrate distribution.  The solid curve represents the best fit
of the $\rho_{\rm QSO}(\theta)$ to the sum of all 5 bands.  The shape
parameters of the dotted and dashed curves have been determined separately using
countrates in three soft and two hard energy bands, respectively.  The profiles
shown in Fig.~\ref{qso_psf} have been obtained by fitting two remaining
parameters, $\rho_0$ and $\rho_b$. Shapes of the PSF fitted to the soft and
hard bands are practically indistinguishable and in the subsequent calculations
we have used the PSF shape parameters represented by the solid curve.

\subsection{The surface brightness distribution \label{brightness}}

At separations comparable to the width of the PSF, i.e. below $\sim\!1\arcmin$,
the countrates generated in the $\delta\rho_i(\theta)$ by the extended emission
are superimposed on the relatively strong signal produced by the point-like source
identified with the galaxy.

\begin{figure}
\centering
\includegraphics[width=9.5cm]{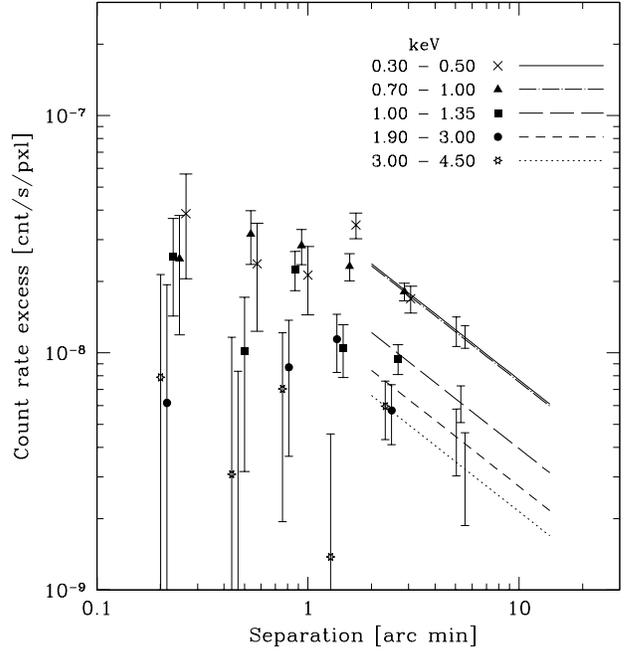}
 \caption{The average excess countrate distribution around galaxies in five energy
bands. For separations above $2\arcmin$ the power law with a slope of $-0.7$ is
assumed and only the normalization has been fitted to the data. At separations
below $2\farcm7$ individual points with the error bars or upper limits are shown.
Some points have been slightly displaced for clarity.}
 \label{rho_ext}
\end{figure}

To subtract the  point source contribution from the $\delta\rho_i(\theta)$
countrates around galaxies, the PSF has been normalized to the observed
distribution at separations below $10\arcsec$ (first three data points in
Fig.~\ref{qso_psf}). The residual flux between $0\fdg2$ and $2\fdg7$ is plotted
in Fig.~\ref{rho_ext}.  The $0.3-0.5$\,keV band is shown with crosses,
$0.7-1.0$\,keV -- triangles, $1.0-1.35$\,keV -- squares, $1.9-3.0$\,keV --
full dots, and $3.0-4.5$\,keV -- with stars. Error bars and upper limits
represent statistical uncertainties at $1\sigma$ level.

At large angular separation -- beyond the WHIM halo surrounding individual
galaxy, enhanced emission results purely from the clustering of galaxies.
According to Eq.~\ref{delta_rho_1}, amplitude of the correlated signal is
defined  by the galaxy ACF. Although the angular extension of a typical halo
around galaxy from the sample is a priori unknown, simulations indicate that
the WHIM signal is negligible at separations above a few arcmin (e.g.
\citealt{bryan01}, \citealt{croft01}). Consequently we tentatively assume that above
$2\arcmin$ the amplitude of $\delta\rho_i(\theta)$ is dominated by the galaxy
correlation term.  The ACF of the APM galaxies has been determined by
\cite{maddox96}:

\begin{equation}
w_{gg}(\theta) = w_0 \cdot \theta^{-0.7},
\label{acf_gg}
\end{equation}
where $w_0 = 0.50$ for $\theta$ in arcmin. Thus, within the framework of
the present model, the count rate distributions $\delta\rho_i(\theta)$ 
at separations greater than $2\arcmin$ has been approximated by the power law:

\begin{equation}
\delta\rho_i(\theta) = A_i \cdot \theta^{-0.7},
\label{rho}
\end{equation}
where the amplitude of the correlation signal is related to the average
integral flux associated with the galaxy:
\begin{equation}
A_i = s_i\,w_0\,n_g.
\label{total_flux}
\end{equation}

The parameters $A_i$ for 5 energy bands have been fitted to the observational
data points for separations between $2\arcmin$ and $14\arcmin$ and the best
power law fits are shown with straight lines in the Fig.~\ref{rho_ext}. The
error bars indicate $1\sigma$ uncertainties of the $A_i$ estimates generated
only by statistical noise.  Because the $\delta\rho_i(\theta)$ functions are
based on a large number of pointings, the scatter of points discernible in
Figs.~\ref{ccf_soft} and \ref{ccf_hard} is adequately described by the Poisson
statistics of the number of counts contributing to each data point. 

\begin{table}
\caption{X-ray fluxes correlated with galaxies}
\label{table_1}
\begin{tabular}{ccrrr}
\hline\hline
$i$&  Energy    & \multicolumn{3}{c} {$s_i\;\;[10^{-5}\,{\rm cnt/s}]$} \\
  &    [keV]    &      total        &   point   & extended\\
\hline
1 & $0.30-0.50$ & $18.3 \pm  2.6$   &   $1.15$  & $17.2$  \\
2 & $0.70-1.00$ & $18.3 \pm  2.0$   &   $1.41$  & $16.9$  \\
3 & $1.00-1.35$ & $ 9.4 \pm  1.7$   &   $0.92$  & $ 8.5$  \\
4 & $1.90-3.00$ & $ 6.5 \pm  2.1$   &   $0.77$  & $ 5.8$  \\
5 & $3.00-4.50$ & $ 5.1 \pm  2.2$   &   $0.82$  & $ 4.3$  \\     
\hline
\end{tabular}
\end{table}

\section{The WHIM structure  \label{results}}

The total fluxes correlated with the individual galaxy in 5 bands have been
determined using Eq.~\ref{total_flux} and are given in Table~\ref{table_1}.
The average density of the APM galaxies in the magnitude range of $17 < b_J <
20$ $n_g = 0.0944$\,arcmin$^{-2}$ was taken from paper on the APM galaxy counts
by \cite{maddox96}.

In the present model, flux per individual galaxy $s_i$ has been split into
point-like and extended components. The average flux contained in the point
source has been estimated by fitting the PSF to the $\delta\rho_i(\theta)$ at
separations below $10\arcsec$  (i.e. first three points in the
$\delta\rho_1(\theta)$ distribution). The integrated flux of the extended
component was obtained by subtraction of the point source from the integral
signal. Both components of the correlated flux are given in columns 4 and 5 of
Table~\ref{table_1}.  Although, the point-like source is conspicuous in the
correlation distribution, $\delta\rho_i(\theta)$, the data in
Table~\ref{table_1} show that on the average the extended source is
substantially stronger than the point-like source.

One should note that in all the energy bands the detected fluxes ascribed to
the individual galaxy are extremely small.  With the mean exposure time for a
single pointing of $\sim\!13$\,ks, the total average signal per galaxy in the
soft bands exceeds just two counts, while the point source amounts to a small
fraction of one count. Thus, the investigated effects are absolutely below the
detection threshold for a single observation and to achieve a reasonable S/N
ratio, the investigation necessarily had to be based on a very extensive data.

Since the galaxy ACF has roughly constant slope of $-0.7$ down to at least
$0\farcm3$ (\citealt{maddox96}), substantial flattening of the
$\delta\rho_i(\theta)$ distribution below $\sim\!2\arcmin$ apparent in
Fig.~\ref{rho_ext} indicates that a dominant fraction of the correlated flux is
generated indeed in the extended source. Large uncertainties involved in the
present analysis and visualized by the error bars in Fig.~\ref{rho_ext} limit
our analysis to the semi-quantitative, gross estimates of the WHIM parameters.
In the subsequent calculations we make assumptions which inevitably introduce
simplifications into the WHIM modeling. Albeit some assumptions cannot be
verified, the present data do not allow for a more detailed description.

In particular, the WHIM distribution below $2\arcmin-3\arcmin$ is approximated
by a well-defined, smooth halo centered on a galaxy. Consequently, the average
surface brightness of the WHIM emission is equal to the amplitude of the
$\delta\rho_i(\theta)$. Such interpretation of the observations is favoured by
the maps of the soft X-ray emission based on hydrodynamic simulations by
\cite{croft01}, where the patches of the highest X-ray surface brightness form
regular circular areas centered on galaxies. Accordingly, the mean angular
extent of the WHIM halo, $\theta_{\rm halo}$, is estimated using the
relationship:

\begin{equation}
\int_0^{\theta_{\rm halo}} \! 2 \, \pi \, \theta \, \delta \rho_i(\theta) \,
                   d\theta = s_i^{\rm ext},
\label{w_model}
\end{equation}
where $\delta \rho_i(\theta)$ is the excess count rate distribution after the
removal of the central peak produced by the point-like component and $s_i^{\rm
ext}$ denotes the flux of the extended source. In rough agreement with
Fig.~\ref{rho_ext}, we assume that for the small separations the excess
countrate $\delta \rho_i(\theta)$ is flat and in the energy band
$0.3-0.5$\,keV, $\delta \rho_1(\theta) \approx 3 \cdot
10^{-8}$\,cnt\,s$^{-1}$\,cm$^{-2}$.  Substituting the relevant quantities into
Eq.~\ref{w_model}, the implied mean halo radius $\theta_{\rm halo} = 2\farcm8$.
Because of the multi-step procedure involved in the calculation, the formal
error estimates of $\theta_{\rm halo}$ are difficult to assess. It is likely
that our ``best estimate'' of the $\delta \rho_i(\theta)$ amplitude at large
separations has been overestimated due to some contribution of the individual
halo to the power law section of the distribution. The lower value of $s_i^{\rm
ext}$ would imply smaller size of the halo. In our opinion, realistic estimate
of X-ray emitting halo size are between $2\arcmin$ and $2\farcm8$. Calculations
of the baryonic mass of the halo described in the Sec.~\ref{densities} below
account for this uncertainty.

\subsection{Temperature of the WHIM \label{temerature}}

Countrates of the extended source listed in the last column of
Table~\ref{table_1} have been used to estimate the mean temperature of the WHIM
emission. We applied the following procedure. First, a grid of thermal plasma
emission spectra based on the \cite{raymond77} code has been generated for a
wide range of temperatures and metal abundances. The MIDAS/EXSAS software has
been used. The spectra have been calculated for redshift $z = 0.12$ (see below)
and subject to the low energy absorption by cold gas in the Galaxy with the
hydrogen column density of $2.2\cdot 10^{20}$\,cm$^{-2}$ (the average value in
the sample).

Next, the spectra have been convolved with the effective area of the X-ray
telescope/EPIC MOS detector system and the model counts in five energy bands
have been compared with the data.  The best fit with $\chi^2 = 10.1$ for 3 dof
was obtained for $kT = 0.50$\,keV and unrealistically low metal abundances
relative to the ``cosmic abundances'' $\zeta = Z/Z_{\rm cosmic} = 0.01$.  We
conclude that the observed counts are inconsistent with a single temperature
model -- the distribution of countrates in 5 energy bands is too wide to be
fitted by a simple thermal spectrum.

A presence of cluster galaxies in our sample provides a natural explanation of
the discrepancy between the data and the single temperature model. Although,
our galaxy sample has been carefully examined from the point of view of the
cluster contamination and all the known clusters have been removed both from the
galaxy sample and the X-ray maps, it is evident that this procedure is not
effective in elimination of all the clusters in the investigated area.  In
particular, our data unavoidably include some unspecified number of X-ray
clusters with the brightest galaxies above the faint magnitude limit of $m_b =
20$. Since the sensitivity of the present investigation is very high, even a
moderate contamination of the galaxy sample with cluster members could
introduce the measurable signal in all the energy bands.

To subtract contribution of the serendipitous clusters to the extended
emission, a thermal spectrum with $kT = 5$\,keV and metal abundances $\zeta =
0.20$ has been adjusted to two hard energy bands ($1.9-3.0$ and
$3.0-4.5$\,keV). Then, the cluster spectrum has been subtracted from the
remaining 3 soft energy bands and the the fitting procedure to the residual
countrates has been repeated.  Obviously, the temperatures of the new fits are
substantially lower than those obtained for the original data. Fits with
acceptable $\chi^2$ have been obtained for temperatures in the range
$0.20-0.30$\,keV with rather weak constraints on the metallicity. For
temperatures close to $kT = 0.20$\,keV high metallicities are favoured, while
for $kT \goa 0.25$ low metallicities $\zeta \loa 0.30$ give acceptable fits.

\section {Baryon densities \label{densities}}

Our estimates of the size and temperature of the average halo are now used to
assess the characteristic densities and the total baryonic mass of the WHIM. In
the calculations we assume that plasma fills the spherical halo with constant
density and the temperature $kT = 0.25$\,keV. According to Sect.~\ref{results},
the halo radii of $2\arcmin$ and $2\farcm8$ are considered.

Roughly $80$\,\% of the APM galaxies in the mgnitude range $17 \le b_J \le 20$
have redshifts between $0.06$ and $0.22$ with the median of the distribution
$z_{\rm med} = 0.12$. (\citealt{maddox96}). Luminosity and angular diameter
distances defined by this redshift are $D_L = 560$\,Mpc and $D_A = 447$\,Mpc,
respectively; angular size  $\theta_{\rm halo} = 2\arcmin$ corresponds to the
radius $r_{\rm WHIM} = 260\,{\rm kpc} = 8.0\cdot 10^{23}$\,cm ($H_0 =
70$\,km/s/Mpc, $\Omega_m = 0.3$, $\Omega_{\Lambda} = 0.7$ is assumed throughout
the paper). We now substitute the results obtained in the paper and all the
relevant data into standard formulae:

\begin{equation}
V_{\rm halo} = {4\over 3}\,\pi\,r_{\rm halo}^3\,,
\end{equation}

\begin{equation}
L_{\rm halo} = V_{\rm halo}\cdot j_E(T)\,,
\end{equation}

\begin{equation}
L_{\rm halo} = 4\,\pi\,D_L^2\cdot S_{\rm halo}\,,
\end{equation}
where $V_{\rm halo}$, $L_{\rm halo}$ and $S_{\rm halo}$ denote respectively:
the halo volume, luminosity and the corresponding flux at distance $D_L$ from
the halo; $j_E(T)$ is the volume emissivity of plasma at energy $E$ and
temperature $T$. Using the basic formulae for the thermal Bremsstrahlung
(e.g. \citealt{rybicki85}) we get for the $0.3-0.5$\,keV energy band:

\begin{equation}
j_E(T) = 6.38\cdot 10^{-25}\, n_p^2 \; {\rm {erg\over s\, cm^3}}\,,
\label{emissivity}
\end{equation}
where $n_p$ is the hydrogen density in cm$^{-3}$. Eq.~\ref{emissivity} is
valid for the zero metallicity plasma with $Y=0.25$, as both the
simulations and our results favour low $Z$ models. Finally, the countrate
$s_1^{\rm ext} = 17.2$\,cnt/s is expressed in the cgs units using the
conversion factor of $1\,{\rm cnt} = 0.65\cdot 10^{-11}$\,erg/s calculated for
the thermal spectrum with the $kT = 0.25$\,keV. Combining all the numerical data
we get the plasma density in the WHIM halo
$\rho_{\rm halo} = 2.3\cdot 10^{-28}$\,g\,cm$^{-3}$
and the baryonic halo mass $M_{\rm halo} = 6.9\cdot 10^{11}$\,M$_{\sun}$ for
the $\theta_{\rm halo} = 2\arcmin$, while for $\theta_{\rm halo} = 2\farcm8$
we have respectively $\rho_{\rm halo} = 3.9\cdot 10^{-28}$\,g\,cm$^{-3}$ and
$M_{\rm halo} = 4.2\cdot 10^{11}$\,M$_{\sun}$.

\subsection{WHIM contribution to $\Omega_{\rm baryon}$}

\begin{figure}
\centering
\includegraphics[width=9cm]{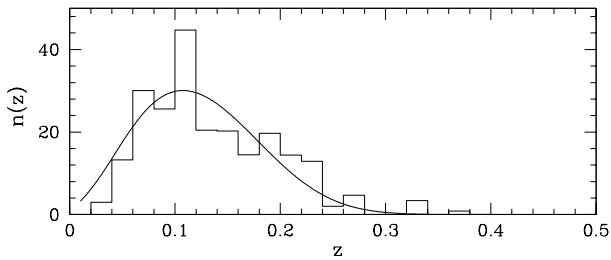}
\caption{The model redshift distribution for galaxies in the magnitude range
$17 < b_J < 20$ derived using the luminosity function given by \cite{folkes99}.
The distribution has been normalized to the histogram of 230 galaxies presented
by \cite{maddox96}.}
\label{z_dist}
\end{figure}

Within the framework of the present model, the WHIM contribution to the
$\Omega_{\rm baryon}$ is defined by the space density of the WHIM halos
centered at galaxies. Because of a large radial density gradient, our magnitude
limited sample of galaxies is not well-suited for such estimates.  To correct
for this effect, the mean spatial density of galaxies contributing to the
observed $\delta\rho_i(\theta)$ distribution averaged over redshift should be
weighted by the number of galaxies in the sample. Let $n(z)$ denotes the
average number of galaxies in the magnitude range $17 < b_J < 20$ as a function
of redshift.  The effective spatial density $\langle N \rangle$ of galaxies in
the sample is given by the formula:

\begin{equation}
\langle N \rangle = {\int N(z)\cdot n(z)\,dz \over \int n(z)\,dz},
\label{space_dens}
\end{equation}
where $N(z)$ is the number of galaxies in $1$\,Mpc$^{-3}$ at redshift $z$
within the apparent magnitude range $17 < b_J < 20$.  Using the luminosity
function of galaxies by \cite{folkes99} adapted to $H_0 = 70$\,km/s/Mpc, we
have determined the both redshift distributions  expected in the APM sample,
$N(z)$ and $n(z)$. The latter one is shown in Fig.~\ref{z_dist} superimposed on
the observed histogram of 230 galaxies in the magnitude range $17 < b_J < 20$
as given by \cite{maddox96}.  After substitution of these distributions into
Eq.~\ref{space_dens} we finally got $\langle N \rangle = 2.4\cdot
10^{-3}$\,Mpc$^{-3}$, and the average WHIM density $\rho_{\rm WHIM} = M_{\rm
halo} / \langle N \rangle = 1.1\cdot 10^{-31}$\,g\,cm$^{-3}$ for $\theta_{\rm
halo} = 2\arcmin$ and $6.8\cdot 10^{-32}$\,g\,cm$^{-3}$ for $\theta_{\rm halo}
= 2\farcm8$.  Assuming the total baryon density $\rho_{\rm baryon} =
0.04\,\rho_{\rm cr} = 3.7\cdot 10^{-31}$\,g\,cm$^{-3}$, the WHIM detected in
the soft X-rays contributes $18 - 31$\,\% to the baryon density, or
$\Omega_{\rm WHIM} = 0.7 - 1.2$\,\%.

\section{Discussion \label{discussion}}

The present investigation is concentrated on two objectives.  First, we
demonstrate an existence of the extended soft emission correlated with
galaxies, which in a natural way is identified with the thermal emission of the
WHIM. Second, using the estimates of the X-ray flux correlated with galaxies,
we determine the physical parameters of the WHIM: the temperature, plasma
density and the total mass of the emitting clouds. Very low amplitude of the
extended emission superimposed on the granular XRB generated by discrete
sources, makes the analysis difficult and sensitive to the cosmic and
instrumental effects.

The first objective was achieved by the careful evaluation and extraction of
several interfering signals and by using an extensive observational material.
The cross-correlation technique appears to be a highly effective tool in the
present analysis. This is because it integrates the delicate X-ray glow
surrounding galaxies and efficiently eliminates various intruding signals. The
large number of pointings and galaxies reduced the fluctuations virtually to
the Poissonian limit defined by the number of galaxy--X-ray count pairs.  Our
estimates of the amplitude of the extended emission in 5 energy bands have been
obtained  under the assumption that the WHIM distribution is correlated with
galaxies, without any further constraints. Strong correlation between the WHIM
and the galaxy distribution is indicated by all the hydrodynamical simulations
(e.g. \citealt{dave01, bryan01, croft01}).

Detailed physical characteristics of the WHIM determined in the paper are not
free from assumptions which -- albeit plausible -- could not be verified by the
observations. Most of our estimates of the plasma parameters have been obtained
under the assumption that the typical WHIM cloud has spherical shape.  The
two-point correlation function gives the WHIM signal averaged over the
azimuthal angle. Thus, the method is unable to identify filaments.  Although
most simulations predict that the WHIM distribution exhibits filamentary
structure, it is likely that the spherical symmetry approximation describes the
WHIM emission satisfactorily. This is because, the elongated structures
``visible'' in the WHIM simulations have extremely low density and surface
brightness, while the regions with high surface brightness are roughly
circular. This effect is visualized in Fig.~1 in \cite{bryan01} and Figs.~3 and
5 in \cite{croft01}.

In the calculations of the halo mass we assumed smooth and uniform distribution
of the gas within a sphere surrounding the galaxy. Such approximation makes our
calculations possible but it certainly constitutes a simplification to the
actual halo structure. Nevertheless, simulations provide some justification for
our approach. A qualitative inspection of the diffuse emission maps generated
by \cite{croft01} reveals surprisingly simple structure at scales typical for
the individual halo. Apparently the simulated structures seem to be fairly well
modeled in the present paper. Although ``visual agreement'' between the maps
and the simple halo model of the WHIM supports our calculations, the range of
the parameter uncertainties of the single halo is probably wider than that
formally assessed in the paper. Consequently, our estimate of the average WHIM
density also are subject to higher uncertainties. A long chain of calculations
constrained by the series of assumptions make our error estimates disputable.
On the other hand, our best estimates of some parameters (i.e. the plasma
temperature and density, contribution of the WHIM to the total baryon density)
fit well to the corresponding quantities obtained in the hydrodynamical
simulations.

One should note that to constrain the simulations, we need substantially more
accurate observational material. It is doubtful that the WHIM emission at
separations above several arcmin from the galaxy could be measured with
substantially better precision using the present day instruments.  However, at
smaller separations, below $2\arcmin - 3\arcmin$, where the WHIM surface
brightness amounts to several percent of the total soft XRB, a new deep
observations of a moderately numerous sample of galaxies at low redshifts would
significantly improve our data. Also, at separations of several arcsec one
could take advantage of the superb angular resolution of {\it Chandra X-Ray
Observatory} and measure the WHIM emission in the immediate vicinity of galaxies.

\begin{acknowledgements}
We thank all the people involved in the XMM-Newton project for making the XMM
Science Archive and Standard Analysis System such user-friendly environments.
This research has made use of the NASA/IPAC Extragalactic Database (NED) which
is operated by the Jet Propulsion Laboratory, California Institute of
Technology, under contract with the National Aeronautics and Space
Administration. This research has also made use of the MAPS Catalog of POSS I
supported by the University of Minnesota.
This work has been partially supported by the Polish KBN grant 1~P03D~003~27.
\end{acknowledgements}


\begin{thebibliography}{}

\bibitem[Alexander et al.\ (2003)]{alexander03}
   Alexander, D. M., Bauer, F. E., Brandt, W. N., et al. 2003, AJ, 126, 539

\bibitem[Bryan \& Voit (2001)]{bryan01}
   Bryan, G. L. \& Voit, G. M. 2001, ApJ, 556, 590

\bibitem[Cen \& Ostriker  (1999)]{cen99}
   Cen, R. \& Ostriker, J. P. 1999, ApJ 514, 1

\bibitem[Croft et al.  (2001)]{croft01}
   Croft, R. A. C., Di Matteo, T., Dav\'e, R., et al. 2001, ApJ, 557,67

\bibitem[Dav\'e et al.\  (2001)]{dave01}
   Dav\'e, R., Cen, R., Ostriker, J. P., et al. 2001, ApJ, 552, 473

\bibitem[Folkes et al.\ (1999)]{folkes99}
   Folkes, S., Ronen, S., Price, I., et al. 1999, MNRAS 308, 459

\bibitem[Kuntz et al. \ (2001)]{kuntz01}
   Kuntz, K. D., Snowden, S. L., \& Mushotzky, R.F. 2001, ApJ 548, L119

\bibitem[Maddox et al.\ (1990b)]{maddox90b}
   Maddox, S. J., Efstathiou, G., \& Sutherland, W. J. 1990, MNRAS, 246, 433

\bibitem[Maddox et al.\ (1996)]{maddox96}
   Maddox, S. J., Efstathiou, G., \& Sutherland, W. J. 1996, MNRAS, 283, 1227

\bibitem[Maddox et al.\ (1990a)]{maddox90a}
   Maddox, S. J., Sutherland, W. J., Efstathiou, G., \& Loveday, J. 1990,
   MNRAS, 243, 692

\bibitem[Nevalainen et al.\ (2005)]{nevalainen05}
   Nevalainen, J., Markevitch, M., and umb, D. 2005, ApJ, 629, 172

\bibitem[Nicastro et al. (2005)]{nicastro05}
   Nicastro, F., Mathur, S., Elvis, M., et al. 2005, Nature 433, 495

\bibitem[Pedersen et al.\ (2005)]{pedersen05}
   Pedersen, K., Rasmussen, J., Sommer-Larsen, J., et al. 2005,
   {\it astro/ph/0511682}

\bibitem[Pradas \& Kerp\ (2005)]{pradas05}
   Pradas, J., \& Kerp. 2005, A\&A, 443, 721

\bibitem[Raymond \& Smith (1977)]{raymond77}
   Raymond, J. C., \& Smith, B. W. 1977, ApJ Suppl. 35, 419

\bibitem[Rybicki \& Lightman (1985)]{rybicki85}
   Rybicki, G. B., \& Lightman, A. P. 1985, {\it Radiative Processes
  in Astrophysics}, New York: Wiley-Interscience

\bibitem[So\l tan et al.\ (2002)]{soltan02}
   So\l tan A. M., Freyberg, M., Hasinger G. 2002, A\&A, 395, 475

\bibitem[So\l tan et al.\ (2005)]{soltan05}
   So\l tan A. M., Freyberg, M., Hasinger G. 2005, A\&A, 436, 67

\bibitem[So\l tan et al.\ (1996)]{soltan96}
   So\l tan A. M., Hasinger G., Egger, R., Snowden, S., \& Tr\"umper, J.
   1996, A\&A, 305, 17

\bibitem[So\l tan et al.\ (1997)]{soltan97}
   So\l tan A. M., Hasinger G., Egger, R., Snowden, S., \& Tr\"umper, J.
   1997, A\&A, 320, 705

\bibitem[Tripp et al. (2000)]{tripp00}
   Tripp, T., Savage, B., \& Jenkins, E. 2000, ApJ, 534, L1

\bibitem[Worsley et al.\ (2005)]{worsley05}
   Worsley, M. A., Fabian, A. C., Baure, F. E., et al. 2005, MNRAS, 357 1281

\bibitem[Zhang \& Pen (2002)]{zhang02}
   Zhang, P. J., \& Pen, U. L. 2002, ApJ, 588, 704

\end{thebibliography}
\end{document}